\documentclass[preprint,superscriptaddress,showpacs,nofootinbib,]{revtex4-1}

\usepackage{amsmath}
\usepackage{slashed}
\usepackage{amssymb}
\usepackage{mathrsfs}
\usepackage[breaklinks]{hyperref}
\usepackage{graphicx}
\usepackage{subfigure}

\usepackage{multirow}

\begin{document}

\title{ Production of charged $\rho$ meson in bottom hadron charmed decays and the effect of the finite width correction of the $\rho$ meson}
\author{Zhen-Hua~Zhang}
 \email[Email:]{zhangzh@iopp.ccnu.edu.cn}
  \affiliation{Institute of Particle Physics, Huazhong Normal University, Wuhan 430079, China}
\author{Ya-Dong~Yang}
 \email[Email:]{yangyd@iopp.ccnu.edu.cn}
  \affiliation{Institute of Particle Physics, Huazhong Normal University, Wuhan 430079, China}
\author{Xin-Heng~Guo}
 \email[Corresponding author, Email: ]{xhguo@bnu.edu.cn}
  \affiliation{College of Nuclear Science and Technology, Beijing Normal University, Beijing 100875, China}%
\author{Gang~L\"{u}}
 \email[Email: ]{ganglv@haut.edu.cn}
  \affiliation{College of Science, Henan University of Technology, Zhengzhou 450001, China.}
\author{Anton~Wiranata}
 \email[Email:]{awiranata@lbl.gov}
  \affiliation{Institute of Particle Physics, Huazhong Normal University, Wuhan 430079, China}
  \affiliation{Lawrence Berkeley National Laboratory, Nuclear Science Division, MS 70R0319, Berkeley, CA 94720, USA}

 
  \date{\today}
  
\begin{abstract}
We calculate the branching ratio of a bottom hadron decaying into a charmed hadron and a charged $\rho$ meson within the QCD factorization approach.
We consider the effect of the finite width correction of the $\rho$ meson.
Our numerical calculation shows an obvious correction because of this effect.
We find that the finite width effect of the $\rho$ meson reduces the branching ratios by about 9\% to 11\% for bottom meson decay channels: $B^+\to\overline{D^0}\rho^+$, $B^0\to D^-\rho^+$, and $B^0_s\to D^-_s\rho^+$,
and increases the branching ratio by about 10\%  for $\Lambda_b^0\to\Lambda_c^+\rho^-$.
\end{abstract}
\pacs{14.20.Mr, 13.30.Eg, 12.39.Hg}

\maketitle

\section{Introduction}
The bottom quark, as one of the third generation of quarks, was very interesting even before its discovery by Fermilab in 1977 \cite{Herb:1977ek}.
The existence of a third generation of quarks is necessary for the complex phase to exist in the Cabibbo-Kobayashi-Maskawa (CKM) matrix, which is the source of CP violation within Standard Model \cite{Kobayashi:1973fv}.
Because of this, studies of the physical processes involving the bottom quark are very important for testing Standard Model and for finding New Physics.
Non-leptonic weak decays of bottom hadrons supply a rich platform to dig the properties of the bottom quark.
The technique of Operator Production Expansion allows us to calculate decay amplitudes of heavy hadrons via an effective Hamiltonian \cite{Buchalla:1995vs}.
In the calculations, one always confront hadronic matrix elements, which contain non-perturbative effects and are hard to handle.
However, with the aid of Heavy Quark Effective Theory \cite{Isgur:1989vq}, one can simplify the hadronic matrix elements via several approaches, such as perturbative QCD \cite{Li:2001vm}, QCD factorization \cite{Beneke:1999br,Beneke:2000ry,Beneke:2001ev,Beneke:2003zv}, and Soft Collinear Effective Theory \cite{Bauer:2000ew,Bauer:2000yr,Bauer:2001ct,Bauer:2001yt}.

The charmed decays of bottom mesons have been studied extensively in  experiments for quite a long time. 
For example,  CLEO reported that the decay branching ratio for $B^0\to D^+\rho^-$ and $B^+\to D^0\rho^+$ are $0.0135 \pm 0.0012 \pm 0.0015$ and $0.0077 \pm 0.0013 \pm 0.0002$, respectively \cite{Alam:1994bi}.

 The properties of bottom baryons are less clear to us than bottom mesons mainly due to its rare production rates compared with bottom mesons.
 However, the decays of $\Lambda_b$ have been becoming more and more clear. 
 The first channel of nonleptonic decay of $\Lambda_b$ yet has been measured is $\Lambda_b\to \Lambda_c \pi$ \cite{Abulencia:2006df}.
 Besides, CDF Collaberation also measured two charmless decay channels of $\Lambda_b$: $\Lambda_b\to p\pi$ and $\Lambda_b\to p K$ \cite{Aaltonen:2008hg}.
 The running of Large Hadron Collider (LHC) gives us a very good opportunity to improve the experimental sensitivity.
 In the near future, LHCb will collect enough events to observe more decay channels of $\Lambda_b$.
 So it is urgent for us to study decays of heavy baryons such as $\Lambda_b$ theoretically.
 
 In this paper, we will focus on one specific kind of decays of the bottom hadrons, in which a bottom hadron decays into a charmed hadron and a charged $\rho$ meson. 
Since the $\rho$ meson decays into two pions strongly, we will take into account its finite width effect.
To our latest knowledge, in most of theoretical considerations for such kind of decays usually a zero width limit for the $\rho$ meson is used. This leads to less accurate results because the decay width of the $\rho$ meson is not small.
 
 This paper is organized as follows. 
 In Sec. \ref{formalism}, we briefly give the formalism for the decay processes we deal with. 
 In Sec. \ref{numericalresult}, we show the numerical results and make some discussions. 
 The conclusion is presented in Sec. \ref{conclusion}.

\section{Formalism \label{formalism}}

Our start point is the effective Hamiltonian for the weak decay $b\to \bar{u}dc$ \cite{Buchalla:1995vs}:
\begin{equation}
\mathscr{H}_{\mathrm{eff}}=\frac{G_F}{\sqrt{2}}V_{ud}^*V_{cb}\left[c_1(\mu)Q_1(\mu)+c_2(\mu)Q_2(\mu)\right]+\text{H.c.},\label{hamilton}
\end{equation}
where $G_F$ is the Fermi constant, $c_1$ and $c_2$ are Wilson coefficients at the scale $\mu$ (which is $\mathcal{O}(m_b)$), $V_{ud}$ and $V_{cb}$ are CKM matrix elements, $Q_1$ and $Q_2$ are four quark operators which take the form
\begin{equation}
\begin{aligned}
Q_1&=\bar{d}\gamma_{\mu}(1-\gamma_5)u\bar{c}\gamma^{\mu}(1-\gamma_5)b, \\
Q_2&=\bar{d}^{i}\gamma_{\mu}(1-\gamma_5)u^{j}\bar{c}^{j}\gamma^{\mu}(1-\gamma_5)b^{i},
\end{aligned}
\end{equation}
where $u$, $d$, $c$, and $b$ represent quark field operators, the superscripts $i$ and $j$ are color indices.
The Wilson coefficients $c_1(\mu)$ and $c_2(\mu)$ take the form \cite{Buchalla:1995vs}
\begin{equation}
\begin{aligned}
c_1(\mu)=\frac{1}{2}\left[z_{+}(\mu)+z_{-}(\mu)\right],\\
c_2(\mu)=\frac{1}{2}\left[z_{+}(\mu)-z_{-}(\mu)\right],
\end{aligned}
\end{equation}
where
\begin{equation}
z_{\pm}(\mu)=\left[1+\frac{\alpha_s(\mu)}{4\pi}B_{\pm}\right]\bar{z}_{\pm}(\mu),\\
\end{equation}
with 
\begin{equation}
\begin{aligned}
\bar{z}_{\pm}(\mu)&=\left[\frac{\alpha_s(m_W)}{\alpha_s(\mu)}\right]^{d_{\pm}}\cdot\left[1+\frac{\alpha_s(m_W)-\alpha_s(\mu)}{4\pi}\left(B_{\pm}-J_{\pm}\right)\right],\\
B_{\pm}&=\frac{N_c\mp 1}{2N_c}(\pm11+\kappa_{\pm}),\\
J_{\pm}&=\frac{d_{\pm}}{\beta_0}\beta_1-\frac{\gamma_{\pm}^{(1)}}{2\beta_0},~~~d_{\pm}=\frac{\gamma_{\pm}^{(0)}}{2\beta_0},\\
\gamma_{\pm}^{(0)}&=6\frac{\pm N_c-1}{N_c},
\gamma_{\pm}^{(1)}=\frac{N_c\mp1}{2N_c}\left[-21\pm\frac{57}{N_c}\mp\frac{19N_c}{3}\pm\frac{4f}{3}-2\beta_0\kappa_{\pm}\right].
\end{aligned}
\end{equation}
In the above equations $m_W$ is the mass of $W$ boson,
$\kappa$ is a scheme dependent parameter, which equals $0(\mp 4)$ for Naive Dimension Regularization ('t Hooft-Veltman) scheme \cite{Buchalla:1995vs}.
Noticing $\bar{z}_{\pm}(\mu)$ is independent of renomalization scheme, we can define scheme independent Wilson coefficients as
\begin{eqnarray}
\bar{c}_1(\mu)=\frac{1}{2}\left[\bar{z}_{+}(\mu)+\bar{z}_{-}(\mu)\right],\\
\bar{c}_2(\mu)=\frac{1}{2}\left[\bar{z}_{+}(\mu)-\bar{z}_{-}(\mu)\right].
\end{eqnarray}

 We will first consider the bottom hadron decay $\Lambda_b^0\to\Lambda_c^+\rho^-$.
 This process was considered in the naive factorization approach long time ago \cite{Acker:1990cq}.
 Here, we will use the QCD factorization approach which was originally constructed for $B$
 meson decays \cite{Beneke:1999br,Beneke:2000ry,Beneke:2001ev,Beneke:2003zv}. 
 When calculating the decay width for $\Lambda_b^0\to\Lambda_c^+\rho^-$, 
after summing over polarizations of $\Lambda_b$ and $\Lambda_c$, one has to typically deal with
\begin{equation}
m_\rho^2\text{tr}\left[\slashed{\epsilon}^{\ast}(\slashed{p}_{\Lambda_b}+m_{\Lambda_b})\slashed{\epsilon}(\slashed{p}_{\Lambda_c}+m_{\Lambda_c})\right],
\end{equation}
where $\epsilon$ is the polarization vector of the $\rho$ meson, $m_\rho$ ($m_{\Lambda_{b(c)}}$) is the mass of $\rho$ ($\Lambda_{b(c)}$), $p_{\Lambda_{b(c)}}$ is the momentum of $\Lambda_{b(c)}$. 
For longitudinally polarized $\rho$ meson ($\lambda=0$), one can replace $m_\rho\epsilon^{(\lambda=0)}$ by $iq$ ($q$ is the momentum of the $\rho$ meson), leading to 
\begin{equation}
m_\rho^2\text{tr}\left[\slashed{\epsilon}^{\ast(\lambda=0)}(\slashed{p}_{\Lambda_b}+m_{\Lambda_b})\slashed{\epsilon}^{(\lambda=0)}(\slashed{p}_{\Lambda_c}+m_{\Lambda_c})\right]\sim m_{\Lambda_b}^4.
\end{equation}
While for transversely polarized $\rho$ meson ($\lambda=\pm$), one has
\begin{equation}
m_\rho^2\text{tr}\left[\slashed{\epsilon}^{\ast(\lambda=\pm)}(\slashed{p}_{\Lambda_b}+m_{\Lambda_b})\slashed{\epsilon}^{(\lambda=\pm)}(\slashed{p}_{\Lambda_c}+m_{\Lambda_c})\right]\sim m_\rho^2m_{\Lambda_b}^2.
\end{equation}
Thus the production of a longitudinally polarized $\rho$ meson dominants. 
 When we consider QCD corrections to the decay $\Lambda_b^0\to\Lambda_c^+\rho^-$, only the vertex corrections contribute in the heavy quark limit \cite{Zhang:2010jc}. 
The decay amplitude for a longitudinally polarized $\rho^-$ in the final state is then
\begin{eqnarray}
\mathcal {A}_{\Lambda_b^0\to\Lambda_c^+\rho^-}
=\frac{G_F}{\sqrt{2}}V_{ud}^*V_{cb}f_{\rho} \langle\Lambda_c|\bar{c}\slashed{q}(a_{1V}-a_{1A}\gamma_5)b|\Lambda_b\rangle, \label{amplitude}
\end{eqnarray}
where
\begin{equation}
a_{1j}=\bar{c}_1(m_b) +\frac{\bar{c}_2(m_b)}{N_c}
\bigg[1+\frac{\alpha_s(\mu)}{4\pi}C_F\int_0^1dx\Phi_{\rho}(x)F_{j}(x,z)\bigg], \label{a_1j}
\end{equation}
here $j=V,A$, $z=m_c/m_b$, $\Phi_{\rho}(x)$ is the light cone distribution amplitude of the $\rho$ meson.
The form of $F_j(z)$ can be found in Ref. \cite{Zhang:2010jc}.

In the heavy quark limit, the hadronic matrix elements $\langle\Lambda_c|\bar{c}\gamma_{\mu}(\gamma_5) b|\Lambda_b\rangle$, which corresponds to the weak transition from $\Lambda_b$ to $\Lambda_c$, can be parameterized as \cite{Isgur:1990pm,Georgi:1990cx,Mannel:1990vg} 
\begin{equation}
\langle\Lambda_c|\bar{c}\gamma_{\mu}(\gamma_5) b|\Lambda_b\rangle=\zeta\left(\omega(s)\right)\bar{u}_{\Lambda_c}(p_{\Lambda_c})\gamma_{\mu}(\gamma_5)u_{\Lambda_b}(p_{\Lambda_b}),
\end{equation}
where $\zeta\left(\omega(s)\right)$ is the Isgur-Wise function \cite{Isgur:1989vq,Isgur:1989ed,Isgur:1990pm}, with 
\begin{equation}
\omega(s)=\frac{m_{\Lambda_b}^2+m_{\Lambda_c}^2-s}{2m_{\Lambda_b}m_{\Lambda_c}},
\end{equation}
where $s=q^2$ ($q=p_{\Lambda_b}-p_{\Lambda_c}$).
Now the nonperturbative effects of strong interaction are fully described by the decay constant of the $\rho$ meson and the Isgur-Wise function.
Then the decay width for $\Lambda_b^0\to\Lambda_c^+\rho^-$ is
\begin{eqnarray}\label{decaywidth}
\Gamma^0_{\Lambda_b^0\to\Lambda_c^+\rho^-}=\frac{G_F^2}{32\pi}|V_{ud}V_{cb}|^2f_{\rho}^2m_{\Lambda_b}^3\left(1-z^2\right)^3|\zeta\left(\omega(m_\rho^2)\right)|^2\left(|a_{1V}|^2+|a_{1A}|^2\right).
\end{eqnarray}
As will be seen later, the above expression for the decay width of $\Lambda_b^0\to\Lambda_c^+\rho^-$ actually corresponds to the situation where the decay width of the $\rho^-$ meson goes to zero. 
However, the vector meson $\rho^-$ has a relatively large decay width (comparing with its mass) because it decays rapidly into two pions.
In addition, the mass of the pion is not small compared with the mass of $\rho^-$ meson.
As a result, we should take the decay width of the $\rho$ meson into account.
This means we have to deal with the decay chain $\Lambda_b^0\to\Lambda_c^+\rho^-\to\Lambda_c^+\pi^0\pi^-$.

The decay amplitude of the decay chain $\Lambda_b^0\to\Lambda_c^+\rho^-\to\Lambda_c^+\pi^0\pi^-$ can be parameterized as
\begin{equation}
M_{\Lambda_b^0\to \Lambda_c^+\rho^-(\to \pi^0\pi^-)}=\frac{g_{\mu\nu}}{s_\rho}M_{\Lambda_b^0\to \Lambda_c^+\rho^-}^{\mu}M_{\rho^-\to\pi^0\pi^-}^{\nu},
\end{equation}
where $s_{\rho}=s-m_{\rho}^2(s)+i\sqrt{s}\Gamma_\rho(s)$, and
\begin{eqnarray}
\begin{aligned}
M_{\Lambda_b\to \Lambda_c\rho^-}^{\mu}&=\frac{G_F}{\sqrt{2}}V_{ud}^{*}V_{cb}\sqrt{s}f_{\rho}\langle \Lambda_c|\bar{c}\gamma^{\mu}(a_{1V}-a_{1A}\gamma_5)b|\Lambda_b\rangle, \\
M^{\mu}_{\rho^-\to\pi^0\pi^-}&=g_{\rho\pi\pi}(q_{\pi^0}-q_{\pi^-})^{\mu},
\end{aligned}
\end{eqnarray}
with $s$ being the invariant mass square of the pion pair, $m_{\rho}^2(s)$ the running mass squared of the $\rho$ meson,  $\Gamma_{\rho}(s)$ the decay width of the $\rho$ meson, $g_{\rho\pi\pi}$ the effective coupling constant among $\rho$ and two pions.
The $s$-dependence of the running mass squared $m_{\rho}^2(s)$ take the form \cite{Gounaris:1968mw,Lichard:2006ky}:
\begin{equation}
m_{\rho}^2(s)=m_{\rho}^2+\frac{2\Gamma_\rho m_\rho^2}{(m_{\rho}^2-4m_\pi^2)^{3/2}}\left\{(s-4m_\pi^2)[h(s)-h(m_\rho^2)]+(m_\rho^2-s)(m_\rho^2-4m_\pi^2)h'(m_\rho^2)\right\},
\end{equation}
with 
\begin{equation}
h(s)=\frac{1}{\pi}\sqrt{\frac{s-4m_\pi^2}{s}}\ln\frac{\sqrt{s}-\sqrt{s-4m_\pi^2}}{2m_\pi},
\end{equation}
for $s>4m_\pi^2$.
The decay width of the $\rho$ meson depends on $s$ via \cite{Gounaris:1968mw,Lichard:2006ky}
\begin{equation}
\Gamma_\rho(s)=\left(\frac{s-4m_\pi^2}{m_\rho^2-4m_\pi^2}\right)^{3/2}\cdot\frac{m_\rho^2}{s}\Gamma_\rho=\left(\frac{1-\frac{4m_\pi^2}{s}}{1-\frac{4m_\pi^2}{m_\rho^2}}\right)^{3/2}\cdot\left(\frac{s}{m_\rho^2}\right)^{1/2}\cdot\Gamma_\rho,
\end{equation}
where $\Gamma_\rho$ is the nominal decay width of the $\rho$ meson, which can be read directly from the particle property list in Particle Data Group.
In general, taking into account the $s$-dependence of the effective coupling constant $g_{\rho\pi\pi}$, $\Gamma_\rho(s)$ should take the from
\begin{equation}
\left(\frac{1-\frac{4m_\pi^2}{s}}{1-\frac{4m_\pi^2}{m_\rho^2}}\right)^{3/2}\cdot\left(\frac{s}{m_\rho^2}\right)^{1/2}\cdot\left[\frac{g_{\rho\pi\pi}(s)}{g_{\rho\pi\pi}}\right]^2\cdot\Gamma_\rho,
\end{equation}
where $g_{\rho\pi\pi}=g_{\rho\pi\pi}(m_\rho^2)$.
Through out this paper, we simply assume that $g_{\rho\pi\pi}$ is independent of $s$.
The coupling constant $g_{\rho\pi\pi}$, corresponding to $\rho$ decaying into two pions,  takes the form
\begin{eqnarray}
g_{\rho\pi\pi}^2=\frac{48\pi m_\rho^{2}\Gamma_{\rho\to2\pi}}{\left(m_\rho^2-4m_\pi^2\right)^{3/2}}=\frac{48\pi}{\left(1-\frac{4m_\pi^2}{m_\rho^2}\right)^{3/2}}\cdot\frac{\Gamma_{\rho\to2\pi}}{m_\rho}.
\end{eqnarray}
Because the $\rho$ meson decays dominantly into two pions, we will simply assume that $\Gamma_{\rho\to2\pi}=\Gamma_{\rho}$.

The differential decay width for $\Lambda_b^0\to\Lambda_c^+\rho^-\to\Lambda_c\pi^0\pi^-$ is then
\begin{equation}\label{partialwidth}
\text{d}\Gamma=\frac{1}{(2\pi)^3}\frac{1}{32m_{\Lambda_b}^3}\frac{1}{2}\sum_{\text{spins}}\left|M_{\Lambda_b^0\to \Lambda_c^+\rho^-(\to \pi^0\pi^-)}\right|^2 \text{d}s\text{d}s',
\end{equation}
with $s'$ being the invariant mass square of  $\Lambda_c$ and $\pi^-$.
Integrating over $s'$, we get (in the heavy quark limit)
\begin{equation}\label{partialwidth2}
\begin{aligned}
\frac{\text{d}\Gamma}{\text{d}s}=
\int_{s'_{\text{min}}}^{s'_{\text{max}}} \text{d}s'\frac{\text{d}\Gamma}{\text{d}s\text{d}s'}
=\frac{\sqrt{s}\Gamma_{\rho}(s)/\pi}{[s-m_{\rho}^2(s)]^2+s\Gamma_\rho^2(s)}\cdot\left[\frac{\zeta(\omega(s))}{\zeta(\omega(m_\rho^2))}\right]^2\cdot\Gamma^0_{\Lambda_b^0\to\Lambda_c^+\rho^-},
\end{aligned}
\end{equation}
where the lower and upper bounds of the integral are
\begin{equation}
\begin{aligned}
&s'_{\text{min}}=\frac{1}{2}\left[m_{\Lambda_b}^2+m_{\Lambda_c}^2-s-(m_{\Lambda_b}^2-m_{\Lambda_c}^2)\sqrt{1-\frac{4m_\pi^2}{s}}\right],\\
&s'_{\text{max}}=\frac{1}{2}\left[m_{\Lambda_b}^2+m_{\Lambda_c}^2-s+(m_{\Lambda_b}^2-m_{\Lambda_c}^2)\sqrt{1-\frac{4m_\pi^2}{s}}\right].
\end{aligned}
\end{equation}
For the total decay width, we have to integrate over $s$.
The bounds for this integral are $4m_{\pi}^2$ and $(m_{\Lambda_b}-m_{\Lambda_c})^2$, respectively.
Although the use of QCD factorization is not appropriate when $s$ is close to the upper bound of the integral, it can be seen from Eq. (\ref{partialwidth2}) that the differential decay width is very small when $s$ is in this region, so the contribution of this part is negligible.
As a result, the decay width for $\Lambda_b^0\to \Lambda_c^+\rho^-\to\Lambda_c^+ \pi^+\pi^-$ can be expressed as
\begin{equation}
\Gamma_{\Lambda_b^0\to \Lambda_c^+\rho^-}=R_{\Lambda_b^0\to \Lambda_c^+\rho^-}\times\Gamma^0_{\Lambda_b^0\to\Lambda_c^+\rho^-},
\end{equation} 
where
\begin{equation}\label{RHeavyQuarkLimit}
R_{\Lambda_b^0\to \Lambda_c^+\rho^-}=\int_{4m_\pi^2}^{(m_{\Lambda_b}-m_{\Lambda_c})^2}\text{d}s\frac{\sqrt{s}\Gamma_{\rho}(s)/\pi}{[s-m_{\rho}^2(s)]^2+s\Gamma_\rho^2(s)}\cdot\left[\frac{\zeta\left(\omega(s)\right)}{\zeta\left(\omega(m_\rho^2)\right)}\right]^2.
\end{equation}
Note that if we let the decay width of the $\rho$ meson goes to zero, we will have the Breit-Wigner form as
\begin{equation}
\lim_{\Gamma_\rho\to0}\frac{\sqrt{s}\Gamma_{\rho}(s)/\pi}{[s-m_{\rho}^2(s)]^2+s\Gamma_\rho^2(s)}=\delta(s-m_\rho^2).
\end{equation}
So we find the decay width becomes just $\Gamma^0$ in this limit.

In deriving Eq. (\ref{RHeavyQuarkLimit}), we have applied the heavy quark limit except the bound of integral over $s$.
However, in order to improve numerical accuracy, we will not use this equation in the following.
Instead, we will use \footnote{Note that in this equation, we neglect a term proportional to $(|a_{1V}|^2-|a_{1A}|^2)/(|a_{1V}|^2+|a_{1A}|^2)$ because it is numerically very small ($\sim10^{-3}$). }
\begin{equation}\label{Rratio}
\begin{aligned}
R_{\Lambda_b^0\to \Lambda_c^+\rho^-}=&\frac{1}{\Gamma^0}\int \text{d}s \text{d}s ' \frac{\text{d}\Gamma}{\text{d}s \text{d}s '}\\
=&\int \text{d}s \text{d}s '\frac{1}{16\pi^2}\frac{s}{(m_{\Lambda_b}^2-m_{\Lambda_c}^2)^3}\left(\frac{\xi(\omega)}{\xi(m_\rho^2)}\right)^2\frac{g_{\rho\pi\pi}^2}{[s-m_\rho^2(s)]^2+s\Gamma_\rho^2(s)}\\
&\cdot\big[(m_{\Lambda_b}^2+m_{\Lambda_c}^2)(s'+s'')-4s's''-2m_\pi^2(m_{\Lambda_b}^2+m_{\Lambda_c}^2)+4m_\pi^4\big].
\end{aligned}
\end{equation}

Similarly, when dealing with bottom meson decays: $B\to D\rho$, if we take the decay width of the $\rho$ meson to be zero, we find that the decay width is 
\begin{equation}
\begin{aligned}
\Gamma_{B\to D\rho}^0
=&\frac{G_F^2}{32\pi}\frac{(m_B+m_D)^5(m_B-m_D)^3}{4 m_B^4m_D}|V_{ud}^*V_{cb}a_1(D\rho)|^2f_{\rho}^2[\xi(\omega(m_\rho^2))]^2.
\end{aligned}
\end{equation}
However, if we take into account the decay width of the $\rho$ meson, again have to deal with the decay chain $B\to D \rho\to D \pi\pi$.
The differential decay width is then
\begin{equation}
\begin{aligned}
\frac{\text{d}\Gamma_{B\to D\rho}}{\text{d}s\text{d}s'}=&\frac{1}{(2\pi)^3}\frac{1}{32m_{B}^3} |M_{B\to D\rho(\to2\pi)}|^2\\
=&\frac{1}{(2\pi)^3}\frac{(m_B+m_D)^2}{256m_{B}^4m_D} G_F^2|V_{ud}V_{cb}|^2sf_{\rho}^2|a_{1V}|^2\left[\xi\left(\omega(s)\right)\right]^2(s''-s')^2\frac{g_{\rho\pi\pi}^2}{|s_\rho|^2}.
\end{aligned}
\end{equation}

Similarly, the decay width is modified to
\begin{equation}
\Gamma_{B\to D\rho}=R_{B\to D\rho}\times\Gamma^0_{B\to D\rho},
\end{equation}
where
\begin{equation}
\begin{aligned}
R_{B\to D\rho}=&\int \text{d}s \text{d}s '\frac{1}{16\pi^2}
\frac{s}{(m_{B}^2-m_{D}^2)^3}
\left(\frac{\xi(\omega)}{\xi(m_\rho^2)}\right)^2
\frac{g_{\rho\pi\pi}^2}{[s-m_\rho^2(s)]^2+s\Gamma_\rho^2(s)}\cdot(s''-s')^2.
\end{aligned}
\end{equation}

\section{Numerical results\label{numericalresult}}

Our input parameters are from Paticle Data Group 2012 \cite{PDG2012}.
With the aid of the renormalization group equation for the running coupling constant, we can get $\Lambda_{\text{QCD}}^{(5)}(\overline{\text{MS}})=231\pm9~\text{MeV}$ at 2-loop order, and $\Lambda_{\text{QCD}}^{(5)}(\overline{\text{MS}})=213\pm8~\text{MeV}$ at 3-loop order, where the uncertainties come from $\alpha_{s}(m_Z)$ and $m_Z$.
Note that at 4-loop order $\Lambda_{\text{QCD}}^{(5)}(\overline{\text{MS}})=213\pm8~\text{MeV}$ \cite{PDG2012}. 
It can be seen that the difference between 3-loop order and 4-loop order results is small.
In fact, we can use either the 2-loop order result or the 3- or 4-loop order result to get the numerical values of the Wilson coefficients, the difference among the obtained results are negligible.
Our numerical results for the scheme independent Wilson coefficients are
\begin{equation}
\bar{c}_1(m_b)=1.146\pm0.003,~\bar{c}_2(m_b)=-0.312\mp0.005,
\end{equation}
where the uncertainties come mainly from the uncertainties of the QCD scale and $m_b$.
The numerical results for the coefficients $a_{1V}$ and $a_{1A}$ are then
\begin{equation}
\begin{aligned}
\text{Re}(a_{1V})&=1.057(2)-0.0064\alpha_1^\rho(\mu)+0.0029\alpha_2^\rho(\mu),\\
\text{Im}(a_{1V})&=0.0223(6)+0.032\alpha_1^\rho(\mu)-0.0015\alpha_2^\rho(\mu),\\
\text{Re}(a_{1A})&=1.056(2)-0.0091\alpha_1^\rho(\mu)+0.0005\alpha_2^\rho(\mu),\\
\text{Im}(a_{1A})&=0.0146(4)+0.028\alpha_1^\rho(\mu)-0.0021\alpha_2^\rho(\mu).
\end{aligned}
\end{equation}
where $\alpha_1^{\rho}$ and $\alpha_2^{\rho}$ are Gegenbauer moments for the $\rho^-$ meson \cite{Beneke:2001ev}.
We only keep the uncertainty of the leading term in the Gegenbauer expansion of the $\rho^-$ meson wave function, which comes from the uncertainties of the QCD scale and the masses of bottom and charm quarks.
We use the following Gegenbauer moments for the $\rho^-$  meson \cite{Beneke:2001ev}:
\begin{equation}
\begin{aligned}
\alpha_1^{\rho}&=0.3\pm0.3,&\alpha_2^{\rho}&=0.1\pm0.3.
\end{aligned}
\end{equation}

We use the decay channel $\rho^0\to e^+e^-$ to extract the decay constant of the $\rho$ meson, $f_{\rho}$.
The decay width for this channel can be expressed as
\begin{equation}
\Gamma_{\rho^0\to e^+e^-}=\frac{2\pi\alpha^2f_\rho^2}{3m_\rho},
\end{equation}
which leads to $f_{\rho}=219\pm1~\text{MeV}$. 

For the Isgur-Wise functions, it can be parameterized as
\begin{equation}
\xi(\omega)=\xi(1)e^{-\rho^2(\omega-1)},
\end{equation}
where $\rho^2$ is the slope parameter, and $\xi(1)=1$ in the heavy quark limit.
One can fit these parameters with certain experimental data.
For $B\to D$ transition, according to Heavy Flavour Average Group (HFAG), these parameters are fitted as $\xi(1)\times|V_{cb}|=(42.64\pm1.53)\times10^{-3}$, $\rho^2=1.186\pm0.054$, with a correlation of 0.829 \cite{Amhis:2012bh}.
For $\Lambda_b\to\Lambda_c$ transition, the uncertainties of the fitted parameters are quite large. 
For example, the fitted value for the slope parameter is $\rho^2=1.59\pm1.10$ by DELPHI Collaboration \cite{Abdallah:2003gn}.

We can also adopt some models for the Isgur-Wise function.
The first model we will use is the MIT bag model which was proposed by Sadzikowiski and Zalewski \cite{Sadzikowski:1993iv}.
In this model, the Isgur-Wise function takes the form
\begin{equation}
\xi(\omega)=\left(\frac{2}{1+\omega}\right)^{a+\frac{b}{\omega}},
\end{equation}
where parameters ($a$, $b$) take the values (2, 0.6), (2.7, 0.6) and (3.5,1.2) for $B\to D$, $B_s\to D_s$ and $\Lambda_b\to\Lambda_c$ transitions, respectively. 
For $\Lambda_b\to \Lambda_c$ transition, we use a so-called soliton model proposed by Jenkins, Manohar, and Wise \cite{Jenkins:1992se}, in which $\zeta(\omega)=0.99\mathrm{e}^{-1.3(\omega-1)}$. 

The numerical results are shown is Table \ref{table}.
We can see that the finite width effect of the $\rho$ meson reduces all the branching ratios of the three decay channels of bottom mesons by about $10\%$.  
On the other hand, the finite width effect of the $\rho$ meson enlarges the branching ratio for $\Lambda_b^0\to\Lambda_c^+\rho^-$ by about $10\%$.
The ratio $R$ is  insensitive to Isgur-Wise function models.
The branching ratios depend on Isgur-Wise models more sensitively.
We can also see from Table \ref{table} that for $B^0\to D^-\rho^+$ and $B_s\to D^-_s\rho^+$ the finite width effect of the $\rho$ meson makes the branching ratios closer to the center value of the experimental data while for $B^+\to\overline{D^0}\rho^+$ the situation becomes worse.
In our calculations we worked in the heavy quark limit while dealing with the warm transition matrix elements.
The order-$1/m_b$ corrections may numerically lead to about $10\%$ change for the decay widths.
This is beyond the scope of our present work.

\begin{table}[!hbp]
\caption{\label{table}Branching ratios for different decay modes. IW stands for Isgur-Wise function.}
\begin{tabular}{cccccc}
\hline
Modes & IW  & R & $\mathcal{B}^0 (\%)$ &$\mathcal{B}=R\mathcal{B}^0 (\%)$& exp. $\mathcal{B}$ (\%)  \\
\hline
\multirow{3}{*}{$\Lambda_b\to\Lambda_c\rho$} &MIT bag & 1.11 &1.02 & 1.13 &\multirow{3}{*}{$--$}\\
\cline{2-5}
& soliton & 1.07  &1.74 & 1.86 &\\
\cline{2-5}
& DELPHI & 1.10(7) & 0.57-3.44 &0.67-3.55&\\
\hline
\multirow{2}{*}{$B^+\to\overline{D^0}\rho^+$} & MIT bag &  0.89& $1.13$& 1.00 &\multirow{2}{*}{$1.34\pm0.18$ \cite{Alam:1994bi}}\\
\cline{2-5}
& HFAG  & 0.90(1) &$1.05(6)$& 0.95(5) &\\
\hline
\multirow{2}{*}{$B^0\to D^-\rho^+$} &MIT bag& 0.89 & 1.04 & 0.92 &\multirow{2}{*}{$0.78\pm0.13$ \cite{Alam:1994bi}}\\
\cline{2-5}
& HFAG & 0.90(1) & 0.97(5)& 0.88(5) &\\
\hline
\multirow{1}{*}{$B^0_s\to D^-_s\rho^+$} & MIT bag  &0.89& 0.78 & 0.70 &\multirow{1}{*}{$0.74\pm0.14$ \cite{Louvot:2010rd}}\\
\hline
\end{tabular}
\end{table}

\section{conclusion\label{conclusion}}
In this paper, we calculate the decay widths of the decay channels $\Lambda_b\to\Lambda_c\rho$ and $B\to D\rho$.
We consider the QCD corrections to these decay channels.
In the heavy quark limit, only vertex corrections contribute, and the production of the longitudinally polarized $\rho$ meson dominants.
In the heavy quark limit, the noperturbative effects of strong interactions are fully described by the decay constant of the $\rho$ meson and a form factor corresponding to the weak transition $\Lambda_b\to\Lambda_c$ or $B\to D$.

Since the $\rho$ meson decays dominantly into two pions through strong interaction, it has a very broad decay width comparing with the mass of the $\rho$ meson.
Also, the mass of the pion is comparable with the mass of the $\rho$ meson.
Even in the heavy quark limit, the two ratios $\Gamma_\rho/m_\rho$ and $m_\pi/m_\rho$ do not go to zero.
As a result, we do not neglect the mass of the pion meson in the calculation.

The correction of the finite decay width effect is described by the parameter $R$ in this paper, which is the ratio of the decay width with and without the finite width effect of the $\rho$ meson.
We find that $R=0.89-0.91$ for $B\to D \rho$, indicating a decrease of the branching ratio by about 9\% to 11\%.
For $\Lambda_b\to\Lambda_c\rho$ channel, we find that the branching ratio is increased by about 10\%.

\begin{acknowledgments}
This work was partially supported by National Natural Science Foundation of China under contract Nos. 10975018, 11047166, 11147003, 11175020, 11225523, 11275025, and the Fundamental Research Funds for the Central Universities in China.
\end{acknowledgments}

\bibliography{zzh}

\end{document}